\documentclass[aps,prb,floatfix,showpacs,twocolumn]{revtex4-1}
\usepackage{amsmath}
\usepackage{graphicx}
\usepackage{latexsym}
\usepackage{bbold}
\usepackage{accents}
\usepackage{bm}
\usepackage{color}
\usepackage{enumitem}

\DeclareMathOperator{\Tr}{Tr}
\begin{document}

\title{Extension of Second-Principles Density Functional Theory into the time domain}

\author{ Toraya Fern\'{a}ndez-Ruiz,$^{1}$
         Jorge \'I\~niguez,$^{2,3}$ and 
         Javier Junquera$^{1}$ and 
         Pablo Garc\'{\i}a-Fern\'andez,$^{1}$ }

\affiliation{$^{1}$Departamento de Ciencias de la Tierra y F\'{\i}sica
  de la Materia Condensada, Universidad de Cantabria, Cantabria Campus
  Internacional, Avenida de los Castros s/n, 39005 Santander, Spain
  \\ $^{2}$  Luxembourg Institute of Science and Technology, 
  Avenue des Hauts-Fourneaux 5, L-4362 Esch/Alzette, Luxembourg
  \\ $^{3}$Department of Physics and Materials Science, University of Luxembourg, 
  41 Rue du Brill, L-4422 Belvaux, Luxembourg}

\date{\today}

\begin{abstract}
We present an extension of the second-principles density functional theory (SPDFT) method to perform time-dependent simulations. 
Our approach, which calculates the evolution of the density matrix in real time and real space using the Liouville-von Neumann equation of motion, allows determining optical and transport properties for very large systems, involving tens of thousands of atoms, using very modest computational platforms. 
In contrast with other methods, we show that SPDFT can be applied to a wide variety of materials including both metals and insulators. In particular, we illustrate its capabilities by obtaining the spectra of SrTiO$_3$, diamond and metallic lithium. 
We find that, while SPDFT results in SrTiO$_3$ are quite similar to those obtained from DFT using linear perturbation theory, we observe significant improvements over this method in both diamond and metallic lithium.
The inclusion of electron-electron interactions during the evolution of the density matrix in diamond allows the spectra to more closely resemble those obtained with the Bethe-Salpeter equation than from perturbation theory. 
In lithium time-dependent SPDFT not only predicts interband transitions but also the Drude peak, opening the possibility of detailed ab initio studies of transport properties beyond many of the usual approximations. 
\end{abstract}


\maketitle

\section{Introduction}

During the last decades research on the effect of scale in the properties of molecular 
and solid-state systems has lead to many important discoveries, both at the fundamental 
and applied levels. 
One of the most illustrative examples is related to what happens when materials are reduced 
to the nanometer range, as in those conditions their properties differ significantly from those 
in bulk. 
For instance, silver or gold metal nanoparticles  with a small radius, r$\approx$10-100 nm,
($\sim 10^4-10^7$ atoms), display surface plasmonic resonances that determine an important part of their optical properties.\cite{kravets_cr18,huang_jar10,natsuki_ijmsa15}
These resonances are very sensitive to the material, shape and size of the nanoparticle
and so they display many non-trivial, difficult to rationalize behaviors.
There are many other fields of current, intense investigation where 
length-scale is also important, like ferroelectric domains,\cite{meier_nrm22} material interfaces, 
or topological quasiparticles like skyrmions.\cite{das_nature19,fert_nrm17}
In a previous work\cite{pgf_prb16} we introduced an electronic structure method, called 
{\sl second-principles density functional theory} (SPDFT), 
that allows performing large scale simulations, involving tens or hundreds of thousands of atoms,
which is enough to tackle many of the problems in the nanoscale described above.
Very importantly, this technique makes predictions using material models created
from the output of first-principles simulations (hence the name second-principles)
and does not require the use of experimental data.
It describes all electronic\cite{pgf_prb16} (charge, orbital or magnetic) and lattice\cite{wojdel_jpcm13} 
degrees of freedom, and its accuracy can be systematically improved towards that of 
the small scale DFT calculations used to train the material's model. 
Various applications of the method can be found in the Refs.~\onlinecite{das_nature19,yadav_nature19,ism_jpcc23}.

However, the time variable can also be quite important in many phenomena at
these scales like, for example, the fast-switching phase transition ocurring
in complex materials like VO$_2$,\cite{cavalleri_prl01} the obtention of the relaxation time in 
transport or the de-excitation of hot electrons in a nanoparticle after irradiation.\cite{appavoo_nanolett14}
In fact, many current research efforts are concentrated in manipulating and tuning 
these magnitudes, with an eye on new technological applications.
Disappointingly none of these properties can be directly handled with simple 
density functional theory (DFT) nor {\sl basic} SPDFT.
Moreover, even the introduction of a simple homogenous electric field, 
necessary to describe the behavior of almost every material in electronic devices,
breaks down periodicity in the solid and induces a {\sl non-stationary} state
that is hard to describe using standard methods.\cite{vanderbilt_chapter}
Time-dependent (TD) density functional theory (TD-DFT) was introduced 
to solve these problems and has allowed to gain much insight in, for example,
the optical properties of materials.\cite{runge_prl84,marques_arpc04,adele_ijqc13,casida_arpc12}

In this work we aim to extend SPDFT into the time domain (TD-SPDFT) with the 
goal of using second-principles to simulate the evolution of the electron distribution of 
large systems in out-of-equilibrium conditions, including the presence of homogeneous 
electric fields. 
A secondary objective is to make these simulations with thousand of hundreds of
atoms work on really modest computational platforms, i.e. desktops or 
small (not massively parallel) clusters and for all kinds of systems, like metallic ones,
where large-scale methods have been traditionally difficult to apply.\cite{aarons_jcp16} 
In order to meet these ends, we have avoided a direct implementation of the 
Runge-Gross equations,\cite{runge_prl84,marques_arpc04} opting to solve the Liouville-von Neumann equation\cite{Cohen-Tannoudji,yam_prb03,wang_jcp07} 
in real-space.
Doing so, our method does not use wavefunctions, relying completely on the real-space
density matrix, which allows us to avoid many computational pitfalls.
Moreover, our implementation will also occur in real-time, in contrast with many
linear-response TD-DFT approaches.\cite{casida_arpc12}
The reason behind this choice is the idea of having a general tool that allows us to tackle
a wide variety of phenomena to any order, rather than implementing individual responses of  the system to each perturbation like electric or magnetic fields, atomic movement, etc. at an specific order.
The final goal is having a code that carries out the dynamics of electrons and lattice simultaneously, allowing the study of non-linear effects typical of structural transitions. 
However, in the present work we will focus exclusively on the movement of the electrons.

The main content of the manuscript is given in the theory section (Section \ref{sec:theory}) that, 
in turn is split in three segments:
(i) In subsection \ref{sec:overview} we present a brief overview of stationary SPDFT and
the notation that will be used for the rest of the article;
(ii) In subsection \ref{sec:propagation} we present the method used to propagate the density.
Beyond the basic equation that describes the fluid movement of the electrons using 
a localized basis set, we devote some effort to detail how the periodic, infinite, 
real-space density matrix is written to efficiently use computational resources.
We finally discuss the changes required in this formalism when an homogeneous electric field 
is introduced and how to solve these problems;
(iii) Finally, in subsection \ref{sec:electric-response} we describe how the response of materials to electric fields is taken into account in SPDFT.
This includes the calculation of the current and polarization, as well as a definition of 
oscillating atomic dipoles. 
The manuscript then continues with the technicalities of the calculations (Sec. \ref{sec:computational}) and results section (Sec. \ref{sec:results}) 
that will illustrate both the ability of TD-SPDFT to carry out simulations 
with large number of atoms and to provide accurate predictions.
This will be followed by a final remarks section (Sec. \ref{sec:conclusions}).

\section{Theory}\label{sec:theory}

\subsection{Overview of the method}\label{sec:overview}

In this section we will provide a quick overview of SPDFT with the goal of presenting the key concepts, 
approximations and nomenclature used in the method as a necessary step to introduce the expansion 
into the time domain immediately after. 
SPDFT is based on the idea that most of the electrons do not react strongly when a perturbation
 (like an electric field or change in temperature) is applied to the system.
 Thus, the total electron density is divided into a reference ($n_0$) and a deformation ($\delta n$) contribution,
   \begin{equation}
      n(\vec{r})=n_0(\vec{r})+\delta n(\vec{r}),
      \label{eq:dens_div}
   \end{equation}
where the last term will involve the electrons (and holes) involved in the response of the system.
Here $\delta n(\vec{r})$ is considered as a small perturbation with respect to $n_0(\vec{r})$ that, in non-magnetic insulators, represents the ground state of the system (see Ref. \onlinecite{pgf_prb16}).
This division is then used\cite{pgf_prb16} to expand the DFT energy with $\delta n$ finding that the zeroth order term, $E^{(0)}$, corresponds with the full DFT energy for the reference density.
The corrections to this reference energy only depend on $\delta n$ (and parametrically on $n_0$) which, given its smallness, can be efficiently calculated leading to a fast and accurate approximation of the full DFT energy. 
The expansion is usually taken to second-order,
  \begin{equation}
     E\approx E^{(0)}+E^{(1)}+E^{(2)}+...,
  \end{equation}
resulting in a stationary problem that is equivalent to Hartree-Fock with the important distinction that the interactions are \emph{screened} by the exchange-correlation potential.
In order to retain $\delta n$ small, the application of the method is restricted to problems where atomic bonds are not created or destroyed, i.e. to processes that display an invariant bond topology.

The reference term, $E^{(0)}$, accounts for the energy of the reference density and is valid for any geometry of the lattice. 
In the case of non-magnetic insulators this corresponds to the Born-Oppenheimer ground state energy surface. 
In order to efficiently access its value we employ an accurate expansion of the DFT energy surface using harmonic and anharmonic terms around a reference atomic geometry (RAG).~\cite{wojdel_jpcm13}
Using this concept, the position of the atomic core $\lambda$ is written as follows,
\begin{equation}
\vec{r}_{\bm\lambda} = (\mathbb{1}+\overleftrightarrow{\eta})\vec{\tau}_{\bm\lambda}+\vec{u}_{\bm\lambda},\label{eq:RAG_position}
\end{equation}
where, $\mathbb{1}$ is the identity matrix, $\overleftrightarrow{\eta}$ is an homogeneous strain tensor, $\vec{\tau}_{\bm\lambda}$ is the position of nuclei $\lambda$ in the RAG, $\vec{\tau}_{\bm{\lambda}}$, and $\vec{u}_{\bm\lambda}$ is the absolute displacement of atom $\lambda$ in the cell $\Lambda$.

To describe the electronic wavefunction we make use of geometry-dependent localized Wannier-like functions (WF), $\chi_{\bm{a}}(\vec{r},t)$.
Here the bold subindex, $\bm{a}$, indicates\cite{pgf_prb16} the collection of atomic index ($a$) and cell-vector ($\vec{R}_a$).
For most systems the WFs do not need to expand all the bands but simply concentrate on the more physically relevant ones, typically involving the higher valence and lower conduction bands. 
Assuming that the WF are geometry dependent and that, in turn, the geometry will change in time we can write the total, time-dependent reduced one-electron density,
   \begin{equation}
     \hat{n}(\vec{r},\vec{r}^\prime,t)=\sum_{\bm{a}\bm{b}} 
                     \left\vert \chi_{\bm{a}}(\vec{r},t) \right \rangle 
                     d_{\bm{a}\bm{b}}(t) 
                     \left \langle \chi_{\bm{b}}(\vec{r}^{\, \prime},t) \right \vert,
     \label{eq:densmat}
   \end{equation}
   the reference density,
   \begin{equation}
     \hat{n_0}(\vec{r},\vec{r}^\prime,t)=\sum_{\bm{a}\bm{b}} 
                     \left\vert \chi_{\bm{a}}(\vec{r},t) \right \rangle 
                     d_{\bm{a}\bm{b}}^{(0)}
                     \left \langle \chi_{\bm{b}}(\vec{r}^{\, \prime},t) \right \vert,
     \label{eq:densmat_ref}
   \end{equation}
   and the difference density,
   \begin{equation}
     \delta \hat{n}(\vec{r},\vec{r}^\prime,t)=\sum_{\bm{a}\bm{b}} 
                      \left\vert \chi_{\bm{a}}(\vec{r},t) \right \rangle 
                      D_{\bm{a}\bm{b}}(t) 
                      \left \langle \chi_{\bm{b}}(\vec{r}^{\, \prime},t) \right \vert.
     \label{eq:densmat_diff}
   \end{equation}
Where $d_{\bm{a}\bm{b}}$, $d_{\bm{a}\bm{b}}^{(0)}$ and $D_{\bm{a}\bm{b}}$ are the density matrices corresponding, respectively, to the total, reference and deformation densities. 
Note that, while total and deformation density matrices are time-dependent we will consider the reference density matrix as constant in time. 
The main consequence of the later choice is that, while in non-magnetic insulators Eq.~(\ref{eq:densmat_ref}) describes the ground state exactly, in metals the density matrix
corresponding with the ground state changes with the geometry.
Thus, to describe metals is very important to account for electron-lattice interactions 
since $n_0$ only describes the ground state in the reference geometry.

   The corresponding expression for the total energy is:
   \begin{align}
      E = & E^{(0)} + 
            \sum_{\bm{a}\bm{b}} 
              \left(
                    D_{\bm{a}\bm{b}}^\uparrow+D_{\bm{a}\bm{b}}^\downarrow
              \right)
              \gamma_{\bm{a}\bm{b}} 
      \nonumber \\
      &+ \frac{1}{2}
        \sum_{\bm{a}\bm{b}}
        \sum_{\bm{a}^\prime\bm{b}^\prime} 
      \left\{
           \left(D_{\bm{a}\bm{b}}^\uparrow + D_{\bm{a}\bm{b}}^\downarrow\right)
           \left(D_{\bm{a}^\prime\bm{b}^\prime}^\uparrow + D_{\bm{a}^\prime\bm{b}^\prime}^\downarrow\right)
      \right.
         U_{\bm{a}\bm{b}\bm{a}^\prime\bm{b}^\prime}
      \nonumber \\
      &\left.
          -\left(D_{\bm{a}\bm{b}}^\uparrow - D_{\bm{a}\bm{b}}^\downarrow \right)
           \left(D_{\bm{a}^\prime\bm{b}^\prime}^\uparrow - D_{\bm{a}^\prime\bm{b}^\prime}^\downarrow\right)
         I_{\bm{a}\bm{b}\bm{a}^\prime\bm{b}^\prime}
      \right\}.
   \label{eq:totalenergy1updn}
   \end{align}   
   
So far, SPDFT has been proposed in the collinear approximation, where the one-particle 
Hamiltonian associated to the Hartree-Fock like energy for the $s$ spin-channel 
in the WFs basis is,
   \begin{align}
       h_{\bm{a}\bm{b}}^s = \gamma_{\bm{a}\bm{b}} + 
         \sum_{\bm{a^\prime b}^\prime} &
              \left[
                  \left(
                     D_{\bm{a}^\prime \bm{b}^\prime}^s+
                     D_{\bm{a}^\prime \bm{b}^\prime}^{-s}
                  \right)
                  U_{\bm{a b a^\prime b}^\prime}-
              \right.
              \nonumber \\
         &
              \left.
                  \left(
                     D_{\bm{a}^\prime\bm{b}^\prime}^{s}-
                     D_{\bm{a}^\prime\bm{b}^\prime}^{-s}
                  \right)
                  I_{\bm{ab a^\prime b}^\prime}
              \right].          
       \label{eq:honeR}
   \end{align} 

As can be seen through the previous expressions, given a model for the system its energy 
is completely defined by the density matrix, $d_{\bm{ab}}$ (or, trivially, the deformation density $D_{\bm{a}^\prime\bm{b}^\prime}$)
and some integrals ($\gamma_{\bm{ab}}$, $U_{\bm{ab},\bm{a^\prime b^\prime}}$, $I_{\bm{ab},\bm{a^\prime b^\prime}}$)
obtained from first principles simulations. 
%
%
For the time-dependent extension of SPDFT we will retain the energy and one-electron equations
given in Eqs.~(\ref{eq:totalenergy1updn}) and (\ref{eq:honeR}) and progress from that foundation.

\subsection{The time evolution of the density matrix}\label{sec:propagation}

A molecular or solid-state system out of equilibrium is uniquely characterized by the electron density current and its associated time-dependent electron density. 
This statement is the basis for TDDFT and was rigorously proved by Runge and Gross\cite{runge_prl84}.
They also proposed that the evolution of a multielectron system could be computed using a set of one-electron time-dependent Schr\"odinger equations using an effective potential related to the exchange-correlation functionals of DFT.
Here, instead of following this scheme, we will propagate the one-electron density operator, $\hat{n}$, using the Liouville-von Neumann equation of motion (EOM),\cite{Cohen-Tannoudji}
\begin{equation}
i\hbar \hat{\dot{n}}(t) = \left[\hat{h}(t),\hat{n}(t)\right].
\label{eq:eom-n}
\end{equation} 
There are four main reasons to proceed in this manner;
$(i)$ the density is both an observable and 
the main magnitude employed to calculate the energy in SPDFT, so it seems reasonable to use
it to keep track of the state of the system instead of using the wavefunction;
$(ii)$ Computationally this allows avoiding the calculation of the density matrix 
from the time-dependent wavefunction at each step, which incurs in a significant 
computer load in simulations;
$(iii)$ its use allows working only in real-space, avoiding sampling of reciprocal space and
$(iv)$ similar approches found in the literature, see e.g. Refs.\onlinecite{souza_prb04,zheng_prb07,wang_prb13,yam_prb03,wang_jcp07}, show that linear scaling can be easily implemented under the EOM formulation,\cite{yam_prb03,wang_jcp07} which is ideal for a large-scale method like SPDFT.

In order to find our working equations we need to plug Eq.~(\ref{eq:densmat}) into the EOM, Eq.~(\ref{eq:eom-n}).
In order to carry out the time derivative in the EOM we will consider that the WFs depend on the geometry 
which, in turn, changes with time during, for example, molecular dynamics simulations.
Thus, using the chain rule, we get,
\begin{equation}
\frac{d}{dt} \left\vert \chi_{\bm{a}} \right\rangle 
   = \sum_{\bm{\alpha}} \frac{d\vec{R}_{\bm{\alpha}}}{dt} \cdot \vec{\nabla}_{\vec{R}_{\bm{\alpha}}} \left\vert \chi_{\bm{a}} \right\rangle 
   = \sum_{\bm{\alpha}} \vec{V}_{\bm{\alpha}} \cdot \vec{\nabla}_{\vec{R}_{\bm{\alpha}}} \left\vert \chi_{\bm{a}} \right\rangle \label{eq:WF_tderiv}
\end{equation}
where $\vec{V}_\alpha=d\vec{R}_\alpha/dt$ is the velocity of the atomic core $\alpha$. 
Defining the time-derivative overlap matrix element, $\delta S_{\bm{a}\bm{b}}$ as,
\begin{align}
      \delta S_{\bm{a}\bm{b}}=\left\langle\chi_a\right\vert \frac{d}{dt} \left\vert \chi_b\right\rangle
              =\sum_{\bm{\alpha}} \vec{V}_{\bm{\alpha}} \cdot
               \left\langle\chi_a\right\vert
                  \vec{\nabla}_{\bm{\alpha}}
                \left\vert \chi_b\right\rangle,
\end{align}
and using Eqs. (\ref{eq:densmat}), (\ref{eq:eom-n}) and (\ref{eq:WF_tderiv}) we find (see related, Appendix \ref{sec:time_der}) that the change of the density matrix with time is,
\begin{equation}
\dot{d}_{\bm{a}\bm{b}} = \frac{1}{i\hbar} \left[ h-i\hbar \delta S, d\right]_{\bm{a}\bm{b}}.\label{eq:d-eom}
\end{equation}
As shown in Eq.~(\ref{eq:d-eom}) the density matrix in the basis of the WFs evolves using the Lioville equation with a modified Hamiltonian matrix,
\begin{equation}
h^\prime_{\bm{a}\bm{b}}=h_{\bm{a}\bm{b}}-i\hbar \delta S_{\bm{ab}}
\end{equation}
involving the elements, $\delta S_{ab}$. 
While, formally, this problem is equivalent to the one where the nuclei are fixed (changing $h^\prime_{\bm{ab}}$ by $h_{\bm{ab}}$) 
and the discussion that follows is valid for the general case, 
from this point on we will consider that nuclear velocities are small enough to neglect the last term of the hamiltonian, 
which is proportional to the non-adiabatic coupling term, $ \left\langle\chi_a\right\vert \vec{\nabla}_{\bm{\alpha}} \left\vert \chi_b\right\rangle$.

\subsubsection{Solution of the equation of motion}

Neglecting the hamiltonian dependency on the density matrix, the exact solution of the EOM is,
\begin{equation}
\hat{n}(t_0+\delta t)=e^{-\frac{i}{\hbar} \int_{t_0}^{t_0+\delta t} \hat{h}(t) dt } \hat{n}(t_0) e^{\frac{i}{\hbar} \int_{t_0}^{t_0+\delta t} \hat{h}(t) dt }.
\label{eq:propagation_exact}
\end{equation}
If we further consider that $h_{\bm{ab}}$ is slowly varying in time we have,
\begin{equation}
\hat{n}(t_0+\delta t)=e^{-\frac{i}{\hbar} \hat{h} \delta t } \hat{n}(t_0) e^{\frac{i}{\hbar} \hat{h} \delta t }.
\label{eq:propagation}
\end{equation}
This approximation is valid if $\delta t$ is sufficiently small.

In order to numerically evaluate this solution it is necessary to expand the matrix exponential in a Taylor series.
We have checked that stable, good quality solutions occur when the expansion is to 4$^{th}$ order or above. 
Testing other, alternative solutions for time-propagation proposed in the bibliography, using, for example, the Magnus expansion,\cite{castro_jcp04,gomez_pueyo_jctc18} we have found slower convergence or stability than with Eq.~(\ref{eq:propagation}).

\subsubsection{Implementation in solids and sparcity of the density matrix}\label{sec:densification}

\begin{figure} [h]
    \begin{center}
       \includegraphics[width=1.0\columnwidth]{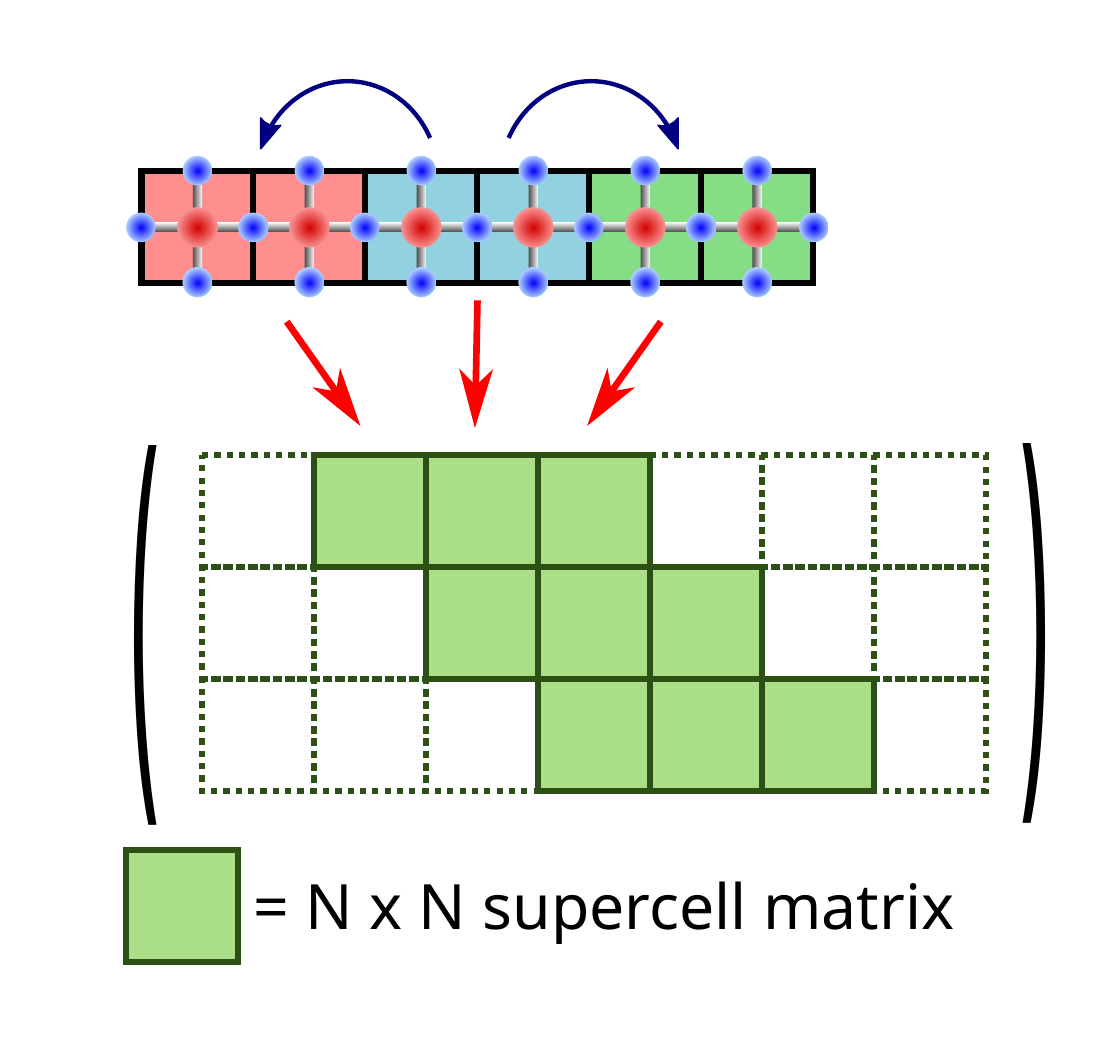}
       \caption{(Color online) Cartoon illustrating the way operators are stored in memory in the code. 
       On the top of the figure we show a linear, periodic system that has been divided in supercells (shown by red, blue and green backgrounds).
       The part of the operator correspond to the N orbitals in each supercell corresponds with a super-row and a super-column in the total operator matrix. 
       The localization of the orbitals allows each of these supercells to interact only with those close to it. Given that the system repeats periodically 
       with each supercell the operator can be simply represented storing just the super-row corresponding to one of the supercells. 
            }
       \label{fig:densification}
    \end{center}
 \end{figure}

The present method uses a pure real space representation, as none of the matrices employed here involve reciprocal space. 
This means that molecular systems can be described using a finite number of basis functions.
However, solids, involving a very large number of crystal cells require the use of an (almost) infinite number of WFs, which poses a challenge to computationally represent the various  operators (hamiltonian and density matrices, etc.).
Here we consider a supercell scheme where the system repeats periodically after the system is translated $N_a$, $N_b$ and $N_c$ conventional cells in the directions of the conventional lattice vectors $\vec{a}$, $\vec{b}$ and $\vec{c}$, respectively.
That means that some properties will be representable using a periodic matrix (see Fig.~\ref{fig:densification}), $A$, that fulfills the following relationship,
\begin{align}\label{eq:periodic}
A_{\bm{a}+\vec{T},\bm{b}+\vec{T}} = A_{a\vec{R}_a+\vec{T},b\vec{R}_b+\vec{T}}=A_{a\vec{R}_a,b\vec{R}_b}=A_{\bm{a},\bm{b}}
\end{align}
where $\vec{T}$ is a translation vector involving an integer number of supercells. 
Using Eq.~(\ref{eq:honeR}) and assuming a periodic density matrix it is trivial to see that the Hamiltonian is also periodic. 
Similarly if we assume an initial density matrix that is periodic in space its evolution using the EOM will also be periodic in space, since the time derivative of the density is also periodic. 
For periodic matrices we do not need to store the information for the whole solid but rather the interaction of a supercell, called the home supercell ($\vec{R}_0$), with the rest of the solid. 
On the other hand other operators, like the position operator,
\begin{equation}
\vec{r}_{\bm{a}\bm{b}}= \left\langle \chi_{\bm{a}} \left\vert \vec{r} \right\vert\chi_{\bm{b}} \right\rangle 
\end{equation}
are non-periodic and cannot be represented in the same way. 

Since the operators are represented in terms of WFs and these are localized and quickly decay with distance,\cite{marzari_rmp12} we expect the Hamiltonian, $h_{\bm{a}\bm{b}}$, and density matrix elements, $d_{\bm{a}\bm{b}}$, to become very small when the WFs $\bm{a}$ and $\bm{b}$ are sufficiently far from each other.
In the case of the Hamiltonian, we observe this behavior when generating the second principles models\cite{pgf_prb16} using {\sc{Wannier90}}\cite{mostofi_cpc08} both for insulators and metals.
Thus, we normally (see the procedure in Ref.~\onlinecite{pgf_prb16} and also Ref.~\onlinecite{carral_arxiv25}) set an energy cutoff for hamiltonian elements that, in turn, restricts the interactions to be within a cutoff-distance, $\delta r_h$.
This allows us to write the hamiltonian as a sparse matrix confined around the origin supercell that can be represented with a finite number of elements. 

In stationary problems the sparsity of the density matrix is the same as that of the hamiltonian. 
However, for non-stationary cases, this is not necessarily so. 
The distance, beyond which we assume that all elements of the density matrix are null, $\delta r_d$, is not defined {\sl a priori}.
For an insulator, the valence density matrix can be expressed as a constant, diagonal matrix,\cite{pgf_prb16} which means that $\delta r_d$ is, essentially, zero. 
However, for metals and other non-trivial cases, like the spectroscopic applications shown later, the density matrix has non-negligible values for off-diagonal elements associated to WFs placed at  finite distance.
Here we take the cutoff-distance for the density, $\delta r_d$, as a computational parameter that we need to optimize for our particular problem.
This is similar to the usual method employed to determine the reciprocal space sampling in periodic boundary DFT calculations, where the density of the integration mesh is gradually increased until the total energy presents small variations and is considered converged.
In our problem, we normally start by setting the density cutoff-distance equal to the hamiltonian cutoff-distance, $\delta r_d= \delta r_h$, which means that all density matrix for WFs separated by less than $\delta r_h$ are considered explicitly in the calculation (i.e. can be non-null).
Then we perform a number of test simulations increasing $\delta r_d$ until the results become stable (normally we check the evolution of the current) and the calculation is considered converged. 

\subsubsection{Application of electric fields}\label{sec:electricfield}

An important goal of the present scheme is to be able to carry out simulations of optical and transport properties in materials, which requires the application of an homogeneous electric field on the system. 
The Hamiltonian matrix element associated to such external electric field is,
\begin{equation}\label{eq:ham_electric}
\hat{h}^E_{\bm{a}\bm{b}} = -e\vec{E}\vec{r}_{\bm{a}\bm{b}}
\end{equation}
Electric fields, break the translational symmetry of the potential in a crystal and force the re-evaluation of the basic blocks of our understanding of solids, like the Bloch theorem.
This can be clearly seen in Eq.~(\ref{eq:ham_electric}) as it depends on the position operator which is non-periodic. 
Fortunately, it has been shown\cite{souza_prb04} that, even with the application of an homogeneous electric field, the evolution of an initially periodic density yields a periodic density at all times. 
This property can be checked, for example, by expanding the exponentials in Eq.~(\ref{eq:propagation}) in a Taylor series when including Eq.~(\ref{eq:ham_electric}) in the hamiltonian.

A computational difficulty that arises here is that, if the hamiltonian is periodic, the evolution operator represented by the exponentials in Eq.~(\ref{eq:propagation}), is also periodic.
This allows storage of this operator as a matrix in a computer. 
However, when including the homogeneous electric field hamiltonian, Eq.~(\ref{eq:ham_electric}), the evolution operator is no longer periodic, preventing the evaluation of Eq.~(\ref{eq:propagation}).
In order to circunvent this problem we notice that the position operator can be separated in a periodic, $\vec{r}^P_{\bm{a}\bm{b}}$, and a non-periodic part,
\begin{align}\label{eq:r_separation}
\vec{r}_{\bm{a}\bm{b}}&= \left\langle \chi_{a\vec{R}_a} \right\vert \vec{r} \left\vert\chi_{b\vec{R}_b}\right\rangle\nonumber\\
&=\left\langle \chi_{a\vec{R}_0} \right\vert \vec{r} \left\vert\chi_{b\vec{R}_b-(\vec{R}_a-\vec{R}_0)}\right\rangle-\delta_{\bm{a}\bm{b}}\left(\vec{R}_a-\vec{R}_0\right)  \nonumber\\
&=\vec{r}^P_{\bm{a}\bm{b}}+\delta_{\bm{a}\bm{b}}\left(\vec{R}_0-\vec{R}_a\right).
\end{align}

The total hamiltonian is the sum of the second-principles, Eq.~(\ref{eq:honeR}), and the electric field contributions, Eq.~(\ref{eq:ham_electric}),
\begin{equation}
h_{\bm{a}\bm{b}}=h_{\bm{a}\bm{b}}^{\text{SP}}+h_{\bm{a}\bm{b}}^{E}
\end{equation}
that can, respectively, be separated into a periodic,
\begin{equation}
h_{\bm{a}\bm{b}}^P=h_{\bm{a}\bm{b}}^{SP}-e\vec{E}\vec{r}^P_{\bm{a}\bm{b}}
\end{equation}
and a non-periodic part,
\begin{equation}
h_{\bm{a}\bm{b}}^{\text{NP}}=-e\vec{E}\left(\vec{R}_0-\vec{R}_a\right) \delta_{\bm{a}\bm{b}}.
\end{equation}
We can now try to decouple\cite{arrighini_ajp96} the non-periodic part of the Hamiltonian from the periodic part in the evolution operator. 
In order to do so we apply the Zassenhaus formula\cite{magnus_cpam54} so that,
\begin{equation}\label{eq:decouple}
e^{-\frac{i}{\hbar} ht} \approx e^{-\frac{i}{\hbar} h^{\text{NP}} t}  e^{-\frac{i}{\hbar} h^P t} e^{-\frac{t^2}{2\hbar^2} [h^{NP},h^P]} \cdots
\end{equation}
In this expression higher terms (the third and remaining elements on the right-hand of Eq.~(\ref{eq:decouple})) involving Poisson brackets are all periodic matrices. 
Thus, the only exponential matrix involving non-periodic exponents is the first one.
The exponent in this exponential matrix is a diagonal matrix and, thus, $e^{-\frac{i}{\hbar} h^{\text{NP}} t}$ can be easily calculated. 

We have checked, as an alternative to the previous scheme, that if the electric field is obtained by choosing a null scalar potential and a vector potential that is linear in time,
\begin{equation}
\vec{E}=-\frac{\partial \vec{A}}{\partial t} \rightarrow \vec{A}=-\vec{E}t
\end{equation}
the whole hamiltonian is periodic but becomes strongly time-dependent. 
Numerically, this introduces significant difficulties evaluating the evolution operator in Eq. (\ref{eq:propagation}), as the exponent becomes larger with time.
Given that the evaluation of exponential matrices relies on Taylor expansions for small exponents, this approach fails for medium-long time evolutions. 

\begin{figure} [b]
    \begin{center}
       \includegraphics[width=1.0\columnwidth]{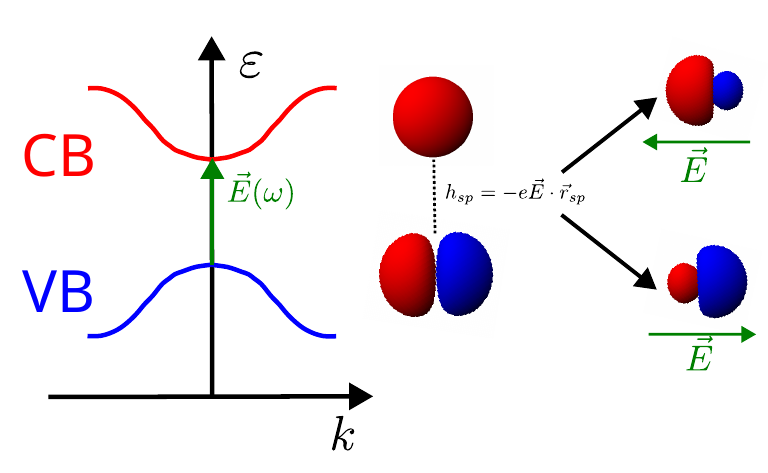}
       \caption{(Color online) Illustration of the effect of the electric field Hamiltonian, Eq.~(\ref{eq:ham_electric}), on the electronic structure of a 
       system that has a valence band formed of WFs with a p$_z$ character and a conduction band with $s$ character. Under the action of the field 
       the two orbitals, that initially do not mix, become entangled 
       creating a hybrid orbital that is polarized to the left or to the right
       depending on the direction of the field.
            }
       \label{fig:sp_hybridization}
    \end{center}
 \end{figure}

In order to carry out simulations that result in the prediction of optical properties
with real-time TDDFT it is quite usual to apply a short, $\delta(t)$, electric-field pulse.
The interaction Hamiltonian that describes this pulse is,
\begin{equation}\label{eq:ham_short_pulse}
\hat{h}^E_{\bm{a}\bm{b}} = -e\vec{E}\delta(t)\vec{r}_{\bm{a}\bm{b}}.
\end{equation}
Introducing this equation in the exact propagation scheme, Eq.~(\ref{eq:propagation_exact}), 
allows us to calculate the density matrix immediately after the application 
of the pulse.
This can be continued with a propagation of the density where no external
electric fields are applied and that allows to capture the response
of the system at all frequencies (since the Fourier transform of the $\delta$-function
in time in Eq.~(\ref{eq:ham_short_pulse}) is a constant in frequency).
Graphically, the effect of this hamiltonian is depicted in Fig.~\ref{fig:sp_hybridization} where it is shown that the effect of an 
electric field is mixing the WFs from the valence band, with p$_z$ character,
and the ones from the conduction band, with $s$ character, producing sp-hybrid 
functions. 

\subsection{Electric response}\label{sec:electric-response}

In this section we will describe how to account in SPDFT for the response of a material
to electric fields.
In the macroscopic formulation of Maxwell's equations the total current, $\vec{J}$,  
the polarization, $\vec{P}$ and the magnetization $\vec{M}$ in a material determine 
its reaction to electric and magnetic fields\cite{jackson_book}. 
In this work we will neglect the effect of the magnetization in the total current
of the system and will focus on the relation between $\vec{J}$ and $\vec{P}$ 
as the gateway to obtain the optical and transport properties of a material, 
given by the complex dielectric tensor, $\overleftrightarrow{\varepsilon}$ and 
the optical conductivity $\overleftrightarrow{\sigma}$. 
Importantly, they are not independent magnitudes as they are related through,
\begin{equation}
\vec{J}=\overleftrightarrow{\sigma} \vec{E} =(\overleftrightarrow{\varepsilon}-\varepsilon_0 \mathbb{1})\frac{\partial \vec{E}}{\partial t}=\frac{\partial\vec{P}}{\partial t}.\label{eq:classcurrpol}
\end{equation}

From the microscopic point of view the polarization in a periodic system was not 
clearly defined until the advent of the {\sl modern theory of 
the polarization}~\cite{resta_chapter,souza_prb04,vanderbilt_prb93,resta_jpcm10,mahon_scipost23} in the 90s. 
Two key aspects of this framework\cite{resta_chapter} are:
(i) the polarization does not take a single, well defined value; 
(ii) in insulators its values form a lattice with a characteristic constant called {\sl quantum of polarization};
(iii) measurements can only capture the change of polarization when a perturbation is applied, not its absolute value.
In order to evaluate the change in polarization the adiabatic current must be integrated 
along the path describing the changes of the system, establishing a connection between 
the current and the polarization through Eq.~(\ref{eq:classcurrpol}).

\subsubsection{The current and the polarization in TD-SPDFT}\label{sec:current}

In SPDFT we will follow a similar path to the one described above: we will first find the current and, from there, 
obtain the polarization, that we will show to be consistent with the points (i)-(iii) raised above.
The total current is the sum of the one created by the nuclei and that associated to the electrons,
\begin{equation}
\vec{J}=\vec{J}_\text{nuc}+\left\langle \vec{J}_\text{elec} \right\rangle.
\end{equation}
This last term that has its usual definition,~\cite{kaxiras_book}
\begin{equation} 
\left\langle \vec{J}_\text{elec} \right\rangle = -\frac{e}{V} \Tr{\hat{n}\hat{\vec{v}}}.
\label{eq:elec_curr}
\end{equation}
In order to express the velocity operator in the WF basis
used in SP-DFT we follow the usual recipe\cite{kaxiras_book,souza_prb04} in which the time derivative of
the position operator is expressed in quantum-mechanical terms.
Using the WF-like basis functions (see Eq.~(\ref{eq_totalder_matrixele}) in App.~\ref{sec:time_der}) we have,
\begin{equation}
    \vec{v}_{\bm{a}\bm{b}}=\dot{\vec{r}}_{\bm{a}\bm{b}} =\frac{i}{\hbar}\left[\hat{h},\hat{\vec{r}}\right]_{\bm{a}\bm{b}} + \frac{\partial^\prime \vec{r}_{\bm{a}\bm{b}}}{\partial^\prime t}, 
\label{eq:vel-wf}
\end{equation}
\noindent where the primed derivative ($\partial^\prime$) indicates the variation of the matrix element only due to the displacement of the basis set functions.

Thus, the current, Eq.~(\ref{eq:elec_curr}), in SP-DFT is,
\begin{align}
\left\langle \vec{J}_\text{elec} \right\rangle 
                      = \frac{e}{V} \sum_{a\bm{c}}\left( 
                   \frac{1}{\hbar}d_{a\bm{c},i}\left[\hat{h},\vec{r}\right]_{\bm{c}a}
                                 -d_{a\bm{c},r}\frac{\partial^\prime \vec{r}_{\bm{c}a}}{\partial^\prime t} 
                                                   \right)                        
\label{eq:current}
\end{align}
where $d_{a\bm{c},i}$ and $d_{a\bm{c},r}$ are, respectively, the real and imaginary parts of the density matrix.
This expression is valid for any system, insulator or metal, working under the assumption 
that the Hamiltonian is real, which is correct in the absence of relativistic corrections 
like spin-orbit.

The relationship of Eq.~(\ref{eq:current}) with the electronic contribution to the polarization
is easily found by comparison with the classical expression, Eq.~(\ref{eq:classcurrpol}),
\begin{align}
\left\langle \vec{J}_\text{elec} \right\rangle 
                      & = \frac{e}{V} \sum_{a\bm{c}}\left( 
                   \frac{1}{\hbar}d_{a\bm{c},i}\left[\hat{h},\vec{r}\right]_{\bm{c}a}
                                 -d_{a\bm{c},r}\frac{\partial \vec{r}_{\bm{c}a}}{\partial t} 
                                                   \right)  \nonumber \\
                                               &=
                        -\frac{e}{V} \sum_{a\bm{c}}\left( 
                                  \dot{d}_{a\bm{c},r}\vec{r}_{\bm{c}a}
                                 +d_{a\bm{c},r}\dot{\vec{r}}_{\bm{c}a}
                                                   \right) \nonumber \\
                  &=\frac{d}{dt}\left(-\frac{e}{V}\sum_{a\bm{c}}d_{a\bm{c}}\vec{r}_{\bm{c}a}\right) \nonumber\\
                  & = \frac{d}{dt}\left\langle \vec{P}_\text{elec} \right\rangle        \label{eq:full_polarization}
\end{align}

We can identity in the above equation the electronic part of the polarization,
\begin{equation}\label{eq:polarization}
\left\langle \vec{P}_\text{elec} \right\rangle=-\frac{e}{V} \Tr{\hat{n}\hat{\vec{r}}}=-\frac{e}{V} \sum_{a\bm{c}} d_{a\bm{c}}\vec{r}_{a\bm{c}}
\end{equation}
\noindent and in order to find the total polarization we would just need 
to add the electron contribution to the nuclear one,
\begin{equation}
\vec{P}=\vec{P}_\text{nuc}+\left\langle\vec{P}_\text{elec}\right\rangle.
\end{equation}

To finish this part of the discussion, we would like to show the relation of the above expression 
with the modern theory of polarization\cite{resta_chapter, vanderbilt_prb93,mahon_scipost23} [see points (i)-(iii) above] by providing a physical interpretation of  Eq.~(\ref{eq:current}).
When the WFs are constructed preventing the mixing of Bloch orbitals coming
from valence and conduction bands, 
the first and second terms of Eq.~(\ref{eq:current}) can be identified 
as the classical free and bound currents, respectively (see Ref.~\onlinecite{carral_arxiv25}). 
In the ground state, the density matrix for an insulator can be chosen 
as a diagonal matrix,
\begin{equation}
d_{\bm{a}\bm{b}}=o_{\bm{a}} \delta_{\bm{a}\bm{b}}
\end{equation}
\noindent where the entry corresponding to a particular WFs, $o_{\bm{a}}$, 
is either 1 or 0.
These values are, respectively, associated to the occupation of
a valence orbital (1 or full) or conduction one (0 or empty).
Thus, the bound current, $\vec{J}_b$, can be written as,
\begin{equation}
\vec{J}_b=-\frac{e}{V}\sum_{a\bm{c}} o_{a}\delta_{a\bm{c}} \frac{\partial \vec{r}_{\bm{c}a}}{\partial t} 
         =\frac{\partial}{\partial t} \left( -\frac{e}{V}\sum_{a} o_{a} \vec{r}_a \right) \label{eq:bound_current}
\end{equation}
and is connected with the movement of the WFs centers.
Thus, the expression inside the parenthesis on the right-hand side of Eq.~(\ref{eq:bound_current})
corresponds with the WF definition of the adiabatic polarization in the modern theory of polarization.\cite{vanderbilt_prb93}
As can be seen in the discussion above, Eq.~(\ref{eq:bound_current}) 
corresponds with an expression valid for insulators and under adiabatic conditions, 
i.e. the Born-Oppenheimer ground state surface. 
On the other hand, the optical polarization, defined in Eq.~(\ref{eq:full_polarization}),
is also valid to be applied under optical conditions and is consistent with the full
current in Eq.~(\ref{eq:classcurrpol}).
It is important to note that Eq.~(\ref{eq:polarization}) is also applicable to metals.
Given that the occupation of the each of the WFs, $o_{\bm{a}}$, in metals  
is a real number between 0 and 1, the absolute value of the polarization is 
neither single valued nor defines a lattice of values, as in the case of insulators 
(the definition of the quantum of polarization relies on occupied WFs having 
the same occupation).
However, once the polarization is assigned a particular value, i.e. choosing a 
particular set of WFs that are placed around sites with position $\vec{r}_{\bm{a}}$, 
Eq.~(\ref{eq:full_polarization}) determines how the optical polarization, $\vec{P}$, changes in time. 
This interpretation is in line with a recent discussion on the application of 
the modern theories of polarization and magnetization to metals.\cite{mahon_scipost23}
It is important to note that there are many ways to define WFs so our definition
of the optical polarization should be considered as an effective way in which 
Eq.~(\ref{eq:classcurrpol}) is fulfilled.

In the original SP-DFT paper\cite{pgf_prb16}, and due to efficiency and physical reasons, 
we separated the one and two electron Hamiltonian elements in short, $h_{\bm{ab}}^\text{sr}$, and 
long range, $h_{\bm{ab}}^\text{lr}$, contributions.
The long range part was defined\cite{pgf_prb16} through the far-field electrostatic potential, $v_{\rm FF}$, created
by a lattice of localized charges and dipoles around each atomic site 
based, respectively, on the total (charges) and reference (dipoles) densities,
and we limited its effect to just a shift of the energy of the orbitals, 
i.e. we took $h_{\bm{ab}}^\text{lr}$ to be diagonal.
While this approach yields good results when treating the low-frequency dielectric response of a system,
it is insufficient to deal with non-stationary situations (optical properties).
As we are interested in following the fast evolution of the polarization with time and the way this screens
external fields it is necessary to take into account the full instantaneous density, Eq.~(\ref{eq:densmat}),
to calculate electrostatic interactions. 
This phenomenon is illustrated in Fig.~\ref{fig:sp_hybridization} where the instantaneous application of a field mixes s and p functions creating 
asymmetric hybrid orbitals that provide a non-null contribution to the polarization.
 
To describe this effect we define the long-range contribution to the Hamiltonian as follows,
\begin{equation}
h_{\bm{ab}}^\text{lr}=-e\left\langle \chi_{\bm{a}} \right\vert v_{\rm FF}[n](\vec{r}) \left\vert \chi_{\bm{b}} \right\rangle
\label{eq:hlong}
\end{equation}
\noindent where the far-field potential is obtained by approximating the electrostatic
integrals around the atomic cores, with position $\vec{r}_{\bm{\lambda}}$ (see Eq.~(\ref{eq:RAG_position})), 
in the way described in our original manuscript.\cite{pgf_prb16} 
The displacements from the RAG of a given atom, that give rise to local dipoles centered around the initial atomic positions, are represented by
\begin{equation}
\delta\vec{r}_{\bm\lambda}= \vec{r}_{\bm\lambda}-\vec{\tau}_{\bm\lambda}= \overleftrightarrow{\eta}\vec{\tau}_\mathbf{\bm\lambda}+\vec{u}_{\bm\lambda}.
\end{equation}
In this case, and taking the full density instead of the reference density, we obtain the same expression
for the electrostatics in terms of atomic contributions,
\begin{align}
v_{\rm FF}[n](\vec{r})=& \sum_{\bm{\lambda}} \frac{Z_{\bm{\lambda}}}{\left\vert \vec{r}-\vec{r}_{\bm{\lambda}} \right\vert}
                         -\int \frac{n(\vec{r}^\prime)}{\left\vert \vec{r}-\vec{r}^\prime \right\vert} d^3r \nonumber\\
                       =& \sum_{\bm{\lambda}} \frac{Z_{\bm{\lambda}}}{\left\vert \vec{r}-\vec{r}_{\bm{\lambda}} \right\vert}
                         -\sum_{\bm{ab}} d_{\bm{ab}} \int \frac{\chi_{\bm{a}}(\vec{r}^\prime)\chi_{\bm{b}}(\vec{r}^\prime)}
                                                              {\left\vert \vec{r}-\vec{r}^\prime \right\vert} d^3r 
                         \nonumber\\
                 \approx& \sum_{\bm{\lambda}} \frac{q_{\bm{\lambda}}}{\left\vert \vec{r}-\vec{\tau}_{\bm{\lambda}} \right\vert}
                         +\sum_{\bm{\lambda}} \vec{p}_{\bm{\lambda}}\frac{\vec{r}-\vec{\tau}_{\bm{\lambda}}}
                                                                         {\left\vert \vec{r} - \vec{\tau}_{\bm{\lambda}} \right\vert^3}
\label{eq:vff}
\end{align}
\noindent which corresponds with a field generated by point charges and dipoles around the atomic positions of the reference atomic geometry (RAG), $\vec{\tau}_{\bm{\lambda}}$.
In the above expression, $Z_{\bm{\lambda}}$, corresponds with the nuclear charge corrected by the core electrons not taken into account
explicitly in the electronic model. 
The changes with respect to the old expressions appear in the definition of the local charges and dipoles which now are:
\begin{align}
q_{\bm{\lambda}}=&Z_{\bm{\lambda}}-\sum_{\bm{a^\prime} \in \bm{\lambda}} d_{\bm{a^\prime a^\prime}} \\
\vec{p}_{\bm{\lambda}}=&Z_{\bm{\lambda}}\delta\vec{r}_{\bm{\lambda}}-\sum_{\bm{a^\prime} \in \bm{\lambda}} 
                                                 \left[
                                                      \sum_{\bm{b^\prime}}\left(
                                                                                d_{\bm{a^\prime b^\prime}}
                                                                                \vec{r}_{\bm{a^\prime b^\prime}}
                                                                          \right) 
                                                                          -d_{\bm{a^\prime a^\prime}} \vec{\tau}_{\bm{\lambda}}
                                                 \right]
\label{eq:atdip}
\end{align}
The second term, in square brackets, is the local dipole generated by the electrons, located at basis function, $\chi_{\bm{a}^\prime}$,
around atom $\bm{\lambda}$. This value involves the difference between the expected contribution of this electron to the polarization 
 and the contribution it would have if it were localized on the atomic position.
Thus, $\vec{p}_{\bm{\lambda}}$ (and $q_{\bm{\lambda}}$) is invariant under translations of a full lattice vector.
It is important to note that compared with our previous approximation\cite{pgf_prb16},
now local dipoles adapt to the instantaneous electronic structure 
that is controlled by both external (applied) and 
internal electric fields (created by the nuclei and electron distribution) and that we will denote, 
respectively, by $\vec{E}_{\text{ext}}$ and $\vec{E}_{\text{int}}$.

In order to calculate the integral in Eq.~(\ref{eq:hlong}) we consider the expansion of the far-field potential, 
Eq.~(\ref{eq:vff}) around the middle point, $\vec{r}_m=\frac{\vec{r}_{\bm{a}}+\vec{r}_{\bm{b}}}{2}$, between the sites of two orbitals $a$ and $b$:
\begin{equation}
v_{\rm FF} (\vec{r},t)\approx v_{\rm FF} \left(\vec{r}_m,t\right) 
                            - \vec{E}_{\text{int}}\left(\vec{r}_m,t\right)\left(\vec{r}-\vec{r}_m\right)
                            +\ldots
\end{equation}
Now calculating the total long-range electrostatic contribution, including internal and external fields, we find,
\begin{align}
h_{\bm{ab}}^{\text{lr}}=&h_{\bm{ab}}^{\text{lr},{\rm FF}}+v_{\bm{ab}}
                       =\left\langle \chi_{\bm{a}} \right\vert 
     -ev_{\rm FF}(\vec{r},t) +e \vec{E}_{\text{total}}(\vec{r},t)\vec{r} 
\left\vert \chi_{\bm{b}} \right\rangle.
\end{align}

Expanding the full long-range electrostatic potential around the point 
$\vec{r}_m$ we get,
\begin{align}
h_{\bm{ab}}^{\text{lr}}\approx&-ev_{\rm FF}\left(\frac{\vec{r}_{\bm{a}}+\vec{r}_{\bm{b}}}{2},t\right)
                               \left\langle \chi_{\bm{a}} \right\vert \left. \chi_{\bm{b}} \right\rangle+\nonumber\\
                              &e\left[\vec{E}_{\text{ext}}(t)
                               +\vec{E}_{\text{int}}\left(\frac{\vec{r}_{\bm{a}}+\vec{r}_{\bm{b}}}{2},t\right)
                                \right]\\
                                &(\vec{r}_{\bm{ab}}-\frac{\vec{r}_{\bm{a}}+\vec{r}_{\bm{b}}}{2}\left\langle \chi_{\bm{a}} \right\vert \left. \chi_{\bm{b}} \right\rangle)\nonumber\\
                             =&-e\left\{v_{\rm FF}\left(\vec{r}_{\bm{a}},t\right)-\left[\vec{E}_{\text{ext}}(t)
                               +\vec{E}_{\text{int}}\left(\vec{r}_{\bm{a}},t\right)\right]\vec{r}_{\bm{a}}\right\}
                               \delta_{\bm{ab}}+\nonumber\\
                              &e\left[\vec{E}_{\text{ext}}(t)
                               +\vec{E}_{\text{int}}\left(\frac{\vec{r}_{\bm{a}}+\vec{r}_{\bm{b}}}{2},t\right)
                                \right]\vec{r}_{\bm{ab}}\nonumber\\
\label{eq:off-diagonal_elec}
\end{align}
Given the expression above we note that the stationary approximation 
was very reasonable given that $\vec{r}_{\bm{ab}}$ elements decay exponentially with the separation 
between $\bm{a}$ and $\bm{b}$. 
Given the necessary closeness between $\bm{a}$ and $\bm{b}$ not to make the off-diagonal matrix element
negligible we can approximate,
\begin{align}
  \vec{E}_{\text{int}}\left(\frac{\vec{r}_{\bm{a}}+\vec{r}_{\bm{b}}}{2},t\right)&\approx
    \frac{1}{2}\left[
                     \vec{E}_{\text{int}}\left(\vec{r}_{\bm{a}},t\right)+
                     \vec{E}_{\text{int}}\left(\vec{r}_{\bm{b}},t\right)
               \right]
\end{align}
\noindent which allows for further computational efficiency. 
Thus, with these changes SPDFT is prepared to describe the change of the charge distribution
in a material as a lattice of charges and dipoles that adapt instantaneously to internal
and external electric fields.

\section{Computational details}\label{sec:computational}

Just like the original stationary Second Principles DFT method, the new extension of to the time domain is implemented in the \textsc{scale-up} (Second-principles Computational Approach for Lattice and Electrons) package \cite{pgf_prb16}, written in Fortran 90 and parallelized using MPI. The parameters of the model needed to obtain the one-electron $E^{(1)}$ two-electron $E^{(2)}$ energies are obtained through a python package called \textsc{modelmaker},\cite{carral_arxiv25} that employs the output of codes using \textsc{wannier90} \cite{pizzi2020wannier90} to create one-electron tight-binding like Hamiltonians. The DFT calculations can be carried out with any code that has interface with \textsc{wannier90} although for this work we have used a direct interface with the \textsc{siesta} \cite{soler2002siesta} code, as it allows using the numerical atomic orbital of the basis set as trial functions to obtain the Hamiltonians, which is very convenient, especially for metallic systems.

In order to illustrate applicability of the method, we have carried out TD-SPDFT calculations in three different systems: (i) The insulating perovskite SrTiO$_3$ that in first approximation can be seen as a ionic material, (ii) diamond, which is a covalent insulating material and (iii) lithium as a prototype metallic solid.

The \textsc{siesta} calculations are perform with the \textit{4.0 version}, considering a 600 Ry mesh cutoff for all the examples treated in the paper. The reference unit cell is taken to be the conventional cell. The reciprocal space is sampled with a Monkhorst Pack $8\times8\times 8$ for SrTiO$_3$ and diamond (insulators), and $15\times15\times 15$ for Li, a bit larger due to its metallic nature. The basis set used to describe the valence electronic states is the default  \textit{double-zeta polarized} (DZP) for the carbon atom in diamond, as it provides an adecuate description of the band structure, a optimized DZP basis set for Li and a optimized \textit{triple-zeta doubly polarized} for SrTiO$_3$. The last two were optimized with the \textsc{simplex} method.\cite{simplex} The kmesh for Wannier considered in the \textsc{wannier90} calculations is $4\times 4 \times 4$ for diamond and SrTiO$_3$ and $8\times 8 \times 8$ for Li. On the other hand, the core states are decribed by a pseudopotential, written in a \textit{.psf} format and optimized following the recipies of the references.\cite{junquera2001numerical,junquera2003first} The optical calculations in \textsc{siesta} are based on first-order perturbation theory, applied by the keyword ``OpticalCalculation''. The sample of the optical (i.e., vertical) transitions along the Brillouin zone is taken $20\times20\times 20$ for the insulators: SrTiO$_3$ and diamond; and $60\times60\times 60$ for lithium.

The scope of one-electron Hamiltonians provided by \textsc{wannier90} are truncated by a cutoff $\delta r_h$ that has been taken as 8\AA \hskip 0.1 cm for all systems. The reason of this selection is because it provides a nice balance between accuracy and computational cost. The trial projections for the Wannier basis set are: (i) 2$p$ oxygen and $3d$-$t_{2g}$ ($d_{xy},d_{yz},d_{xz}$) titanium atomic orbitals for the upper valence and lower conduction bands of SrTiO$_3$. (ii) 2$s$ and 2$p$ atomic orbitals placed in the carbon atoms and (iii) 2$s$ and 2$p$ atomic orbitals placed in the lithium atoms; that give rise to a set of 12, 32 and 8 bands respectively for the conventional cell of each system.

\section{Results}\label{sec:results}

In this section we will discuss the numerical results obtained from the application of TD-SPDFT
to a series of examples, ranging from simple tight-binding models used to test the scalability 
of the code, to more realistic systems like SrTiO$_3$, diamond and metallic lithium to show the 
capabilities of the method.

\subsection{Efficiency test}\label{sec:model}

In order to study the scalability of the implementation of the equations of Section \ref{sec:propagation} 
in {\sc scale-up} we have prepared a simple tight-binding model for a crystal containing 4-atoms
in the unit cell, resembling a cubic ABX$_3$ perovskite without the A-cation. 
The valence bands are formed out of Wannier functions with p-character, localized over the anions (X),
 facing directly towards the B-cation ($\sigma$-bonding), while the latter supports a single s-like Wannier 
 function, that forms the conduction band.
The hamiltonian includes first-neighbor p-p and s-s interactions while the position operator also includes 
s-p elements, to allow for optical transitions between the valence and conduction bands.

  \begin{figure} [b]
    \begin{center}
       \includegraphics[width=1.0\columnwidth]{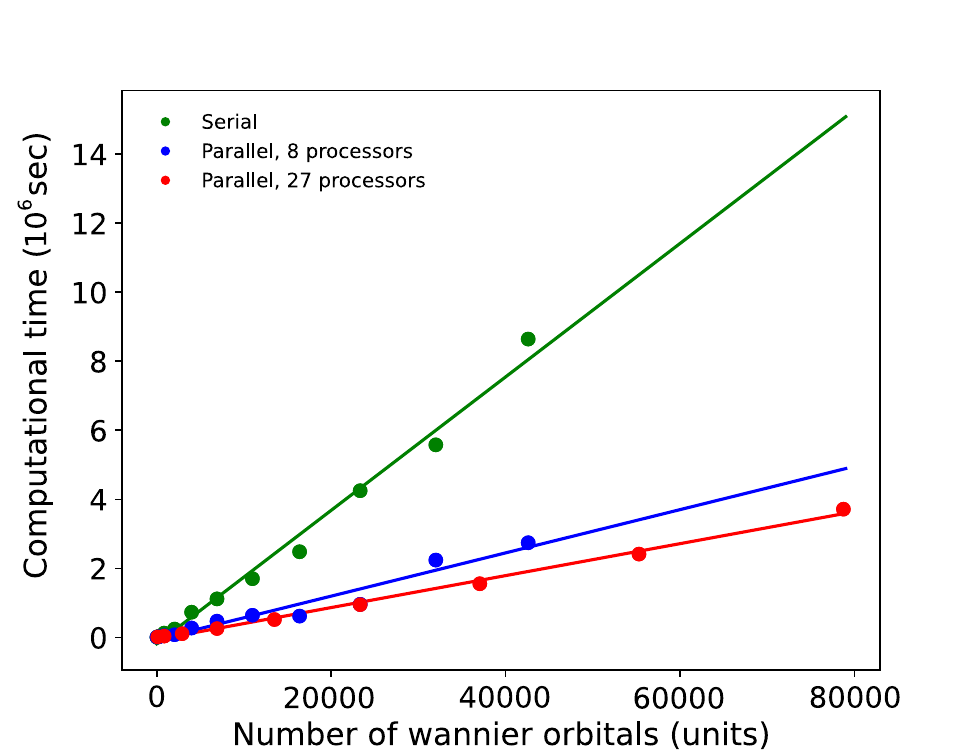}
       \caption{(Color online) Evolution of the computational time with the number of Wannier orbitals in the 
       simulation supercell in the simple tight-binding model described in Sec.~\ref{sec:model} (circles). It can be seen
       that the computational time scales linearly (solid lines) for the serial (green) and parallel runs with 8 (blue) and
       27 processors (red).
            }
       \label{fig:time_linear}
    \end{center}
 \end{figure}
  
Using this setup for the unit-cell, we carry out simulations of the spectrum using the short-pulse 
technique and a total simulation time of 10 fs and a time-step of 0.01 fs.
In Fig.~\ref{fig:time_linear} we represent the total time that takes each simulation using a N$\times$N$\times$N
supercell, when N takes values between 1 (4 atoms) and 27 (78732 atoms), using a single workstation processor.
It can be seen that the computational time grows linearly with the number of Wannier functions in the 
simulations cell, a behavior that is also found in the memory use (not shown).
With the introduction of parallelization (blue and red lines in Fig.~\ref{fig:time_linear}) we retain 
the linearity both in time and memory with a very significant reduction in wall time allowing to reach
much larger number atoms in the simulations cells (typical sizes are tens of thousands of atoms 
with $\approx$30 processors and run times close to a month of wall-time). 
Thus, we have shown that the method is able to carry out simulations with very large number of 
atoms. 
In the following, we will focus on discussion of the quality of the results, whether the system
delivers first-principles-like accuracy, and its ability to treat difficult cases (i.e. metals).
 

\subsection{Optical properties of SrTiO$_3$}

As a first case we will study the optical properties of SrTiO$_3$ an insulator that
presents an experimental gap\cite{vanbenthem_jap01} of 3.25 eV.
We have built a simple tight-binding model for this system using {\sc Wannier90}
over an LDA calculation employing a reciprocal space sampling of 8$\times$8$\times$8
points. 
The Wannier functions for the valence bands were obtained by projecting the Bloch
orbitals over 2p atomic-like orbitals centered on the oxygen atoms. 
The lower part of the conduction band was obtained by a similar projection on 
$t_{2g}$-like ($d_{xy}$, $d_{xz}$, $d_{yz}$) orbitals centered on the titanium ion.
The hamiltonian obtained from this wannierization procedure was then cut with a hamiltonian cutoff $\delta r_h=8.0$ \AA, which is enough to reproduce the bands of the cubic phase as to be indistinguishable from the first-principles ones at simple sight.
We have found that the results were stable with a density cutoff $\delta r_d=8.0$ \AA.
Intraatomic electron-electron interactions\cite{carral_arxiv25} were introduced in this model.

\begin{figure} [h]
    \begin{center}
       \includegraphics[width=1.0\columnwidth]{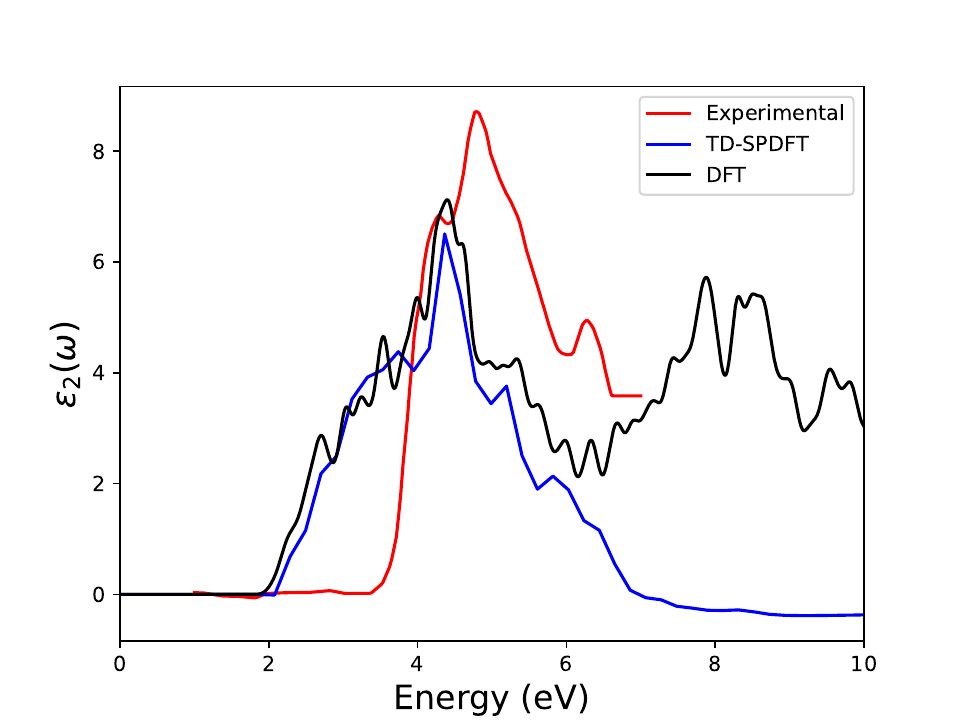}
       \caption{(Color online) Complex-part of the dielectric constant versus frequency for SrTiO$_3$ obtained from 
             first-order perturbation theory in DFT (black lines), TD-SPDFT (blue lines) and
             experimental (obtained from data in Ref.~\onlinecite{vanbenthem_jap01}, red lines). 
             The TD-SPDFT curves have been obtained for $\delta r_d=8.0$ \AA.
            }
       \label{fig:STO}
    \end{center}
 \end{figure}

In order to obtain the spectrum we applied a short-pulse electric field, $E=10^{-4}$ au,
followed by a 20 fs propagation.
The obtained time-dependent current, Eq.~(\ref{eq:current}), was then Fourier transformed into
the frequency realm and the result divided by the intensity of the electric field to obtain
the frequency-dependent conductivity. 
Taking the real part of the conductivity, $\sigma_1$, we obtained the complex part of the 
dielectric function using the expression,
\begin{equation}\label{eq:im_dielectric}
\varepsilon_2 = 4\pi \frac{\sigma_1}{\omega}.
\end{equation}
This result can be compared to the complex part of the dielectric function calculated by
{\sc SIESTA} using perturbation theory and a sum-over-bands approach (see Fig.~\ref{fig:STO}).
As it can be seen, the spectrum obtained using TD-SPDFT is quite similar to the one obtained
with DFT.
The main difference is due to the fact that in the second-principles ones only 
the lower part of the conduction band, with Ti($t_{2g}$) character, is simulated.
At energies larger than 6.0 eV the spectrum has large contributions from other bands [like Ti($e_g$)]
 that are present in first-principles but not in SP.
Moreover, it can be seen that the band gap in both FP and SP is reduced ($\approx$2 eV) when
compared to the experimental one ($\approx 3.25$ eV\cite{vanbenthem_jap01}). 
This can be directly related to the use of LDA in the DFT simulation, which is well-known to  
underestimate the gap. 

We have checked that the same spectrum is obtained independently of the N$\times$N$\times$N  supercell used in the SP simulation. 
We have carried simulations up to N=27 (for a total of 98415 atoms in the unit cell).
This simulation took just over a month using 27 processors.

\subsection{Optical properties of diamond}

Diamond is a wide-gap insulator with an indirect energy gap of 5.48 eV
and a direct gap of 7.3 eV.\cite{saslow_prl66}
In order to simulate its spectrum we created a SP model using {\sc wannier90}
on a DFT calculation carried out with {\sc Siesta} employing PBE
and a standard DZP basis.
This tight-binding model has a cutoff distance, $\delta r_h=8.0$ \AA.
Short-range electron-electron interactions were also introduced in the model. 
It was found\cite{pgf_prb16} that, as expected, intra-atomic electron-electron interactions
were large ($U\approx$4.0 eV) and were added to the model.

  \begin{figure} [b]
    \begin{center}
       \includegraphics[width=1.0\columnwidth]{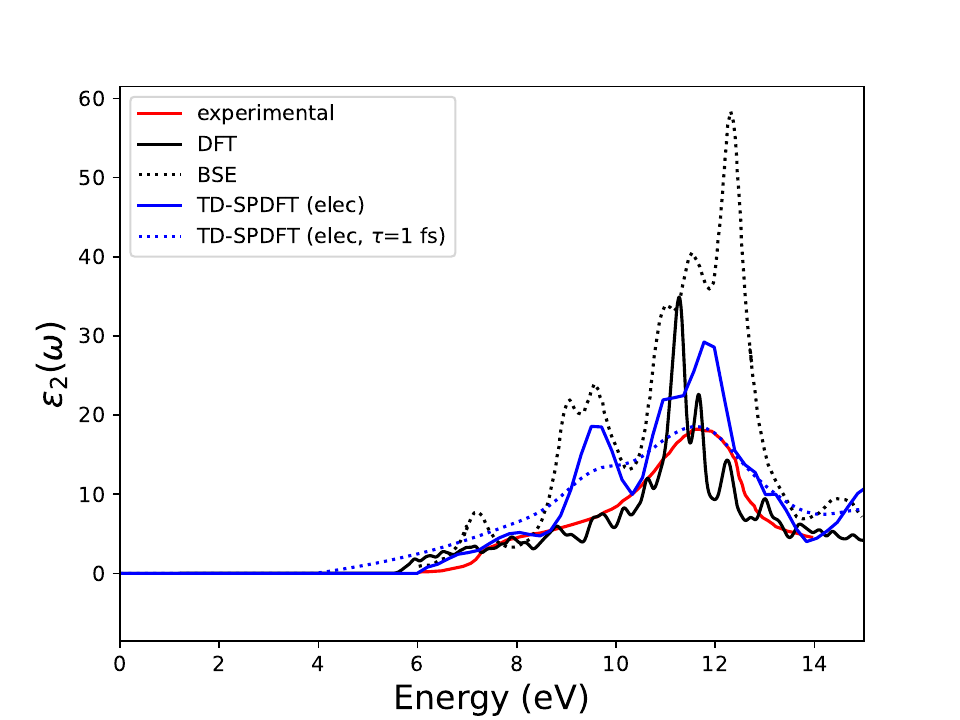}
       \caption{(Color online) Complex-part of the dielectric constant versus frequency for diamond obtained from 
             first-order perturbation theory in DFT (solid black lines), Bethe-Salpeter Equation (BSE, taken from Ref.~\onlinecite{tal_prr20}),
             TD-SPDFT (solid and dotted blue lines, depending on the lifetime used) and
             experimental (obtained from data in Ref.~\onlinecite{palik_optical_book}, red lines). 
             The TD-SPDFT curves have been obtained for $\delta r_d=8.0$ \AA.
            }
       \label{fig:diamond}
    \end{center}
 \end{figure}

As in the case of SrTiO$_3$ the spectrum was obtained by applying a short-pulse electric field 
of $E=10^{-4}$ au intensity, followed by a 20 fs propagation of the density.
After studying the effect of the cutoff radius for the density it was found that the minimal one,
$\delta r_d=\delta r_h=8.0$ \AA,~ produced adequate results and was chosen for all the results 
presented here.
The resulting complex part of the dielectric constant, $\varepsilon_2$, is shown in Fig.~\ref{fig:diamond}.
In that figure the experimental results\cite{palik_optical_book} are first compared to the perturbative results 
obtained from {\sc siesta}.
There we can see that the experimental spectrum, inside the ultraviolet region, is quite broad and featureless.
The perturbation theory result matches reasonably well this spectrum although the main absorption
peak is approximately 1 eV below the experimental maximum. 
This is a reasonable result since DFT usually underestimates excitation energies (gaps). 
If we compare this data with the result of TD-SPDFT calculations (in solid blue) we see that the main
transition is now shifted to higher energies, producing a very nice match with the energy of the
experimental peak. 
However, the TD-SPDFT dielectric constant shows much structure not found in the experimental spectrum.
In particular there is a prominent peak centered around 9.5 eV that does not appear in the experimental measurement.
We have checked that, in order to broaden the simulated spectrum, if the TD-SPDFT time signal (electron current) is convoluted 
with a decaying exponential of lifetime $\tau=1.0$ fs, the resulting curve closely matches the experimental
result. 
Moreover, if the original TD-SPDFT spectrum is compared with Bethe-Salpeter-Equation (BSE) results, obtained from
Ref.~\onlinecite{tal_prr20}, we find that the extra features like the peak at 9.5 eV or the splitting occurring
on the main peak are also present in that much more accurate (and costly) calculation. 
Thus, we find that the TD-SPDFT results are, in fact, quite accurate being able to produce results that 
are not directly found in the DFT data used to generate the model and which is both closely related to
experimental results and higher-level theory. 

\subsection{Metallic lithium}

As a final example of the applications of TD-SPDFT, we approach a more complicated problem involving 
metallic lithium, usually considered as a typical example for quasi-free electrons. 
This is both out of the comfort zone\cite{aarons_jcp16} of simulations involving localized basis 
and linear-scaling methods, which are usually applied in insulators, as well as many theoretical 
approaches to spectroscopy that require finite excitations and dielectric constants.

\begin{figure} [b]
    \begin{center}
       \includegraphics[width=1.0\columnwidth]{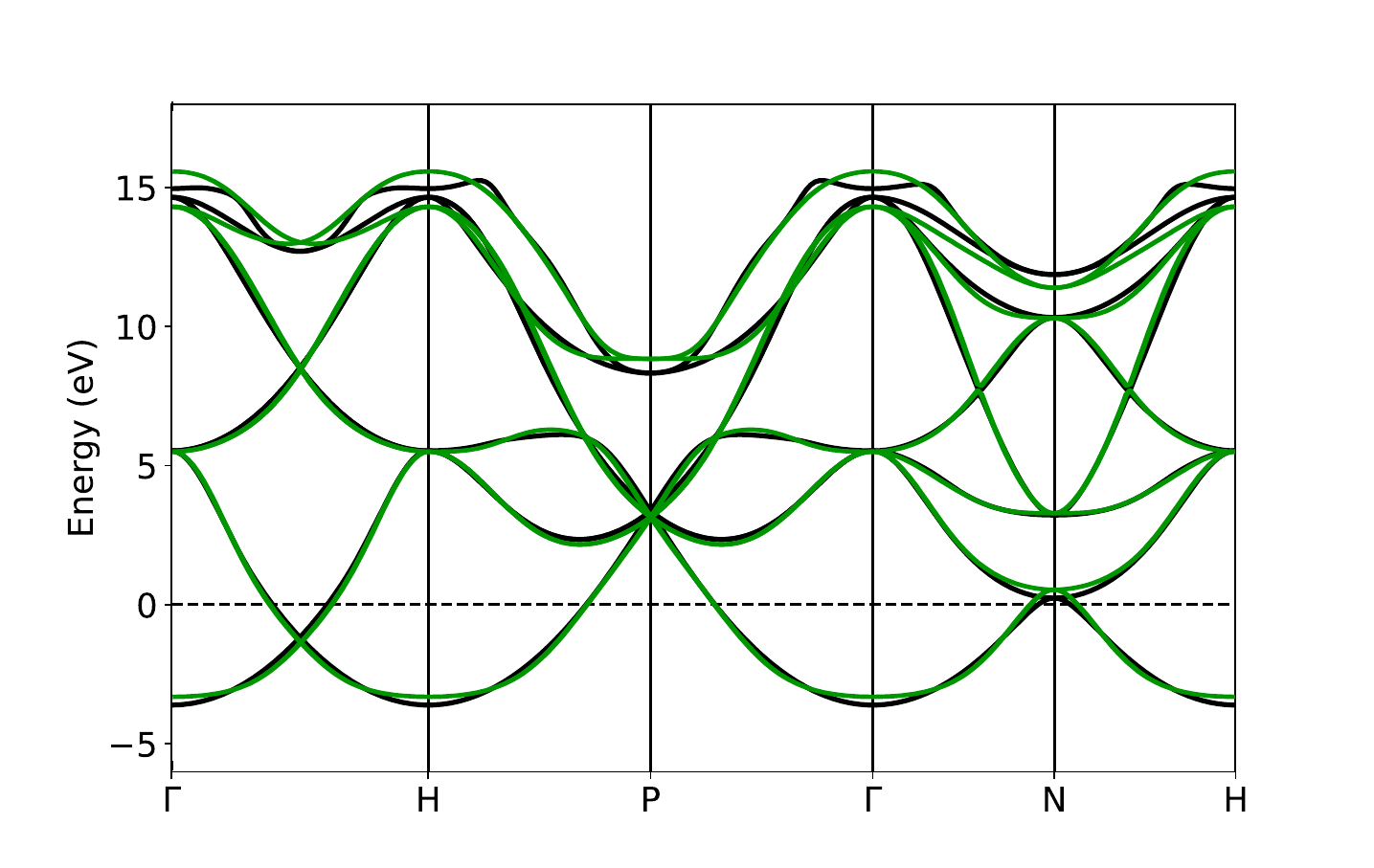}
       \caption{(Color online) Bands for metallic lithium obtained in DFT (black lines) and 
       SPDFT (green lines) using a cutoff $\delta r_h=8.0$ \AA. The Fermi energy is placed at E=0 eV.
            }
       \label{fig:lithium_bands}
    \end{center}
 \end{figure}

We created the SP model for lithium starting from a PBE DFT calculation using {\sc siesta} with an
optimized basis containing two 2s valence orbitals and two sets of 2p polarization orbitals.
The results obtained from this simulation were passed to {\sc Wannier90} to produce 
a tight-binding model including 4 Wannier orbitals per atom by projecting the lower DFT bands and
Bloch orbitals on atomic 2s and 2p orbitals. 
The interaction distance between Wanniers in this hamiltonian, $\delta r_h$, was limited to 8.0 \AA~ which 
allows reproducing the DFT bands with only small discrepancies (see Fig.~\ref{fig:lithium_bands}),
close to the bottom of the conduction band and far from the Fermi energy, 
the most important part for physical properties (E=0 eV in the plot). 

\begin{figure} [b]
    \begin{center}
       \includegraphics[width=1.0\columnwidth]{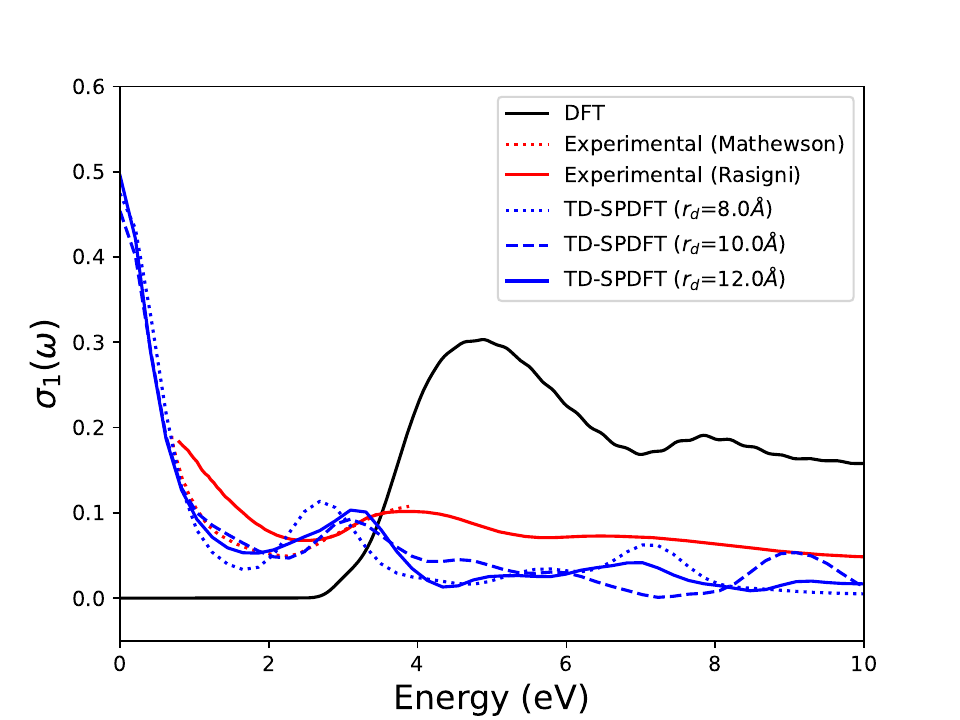}
       \caption{(Color online) Real-part of the optical conductivity for metallic lithium obtained from 
             first-order perturbation theory in DFT (solid black lines) and TD-SPDFT 
             (solid, dashed and dotted blue lines representing, respectively, the curves obtained for $\delta r_d =$8.0, 10.0 and 12.0 \AA with a lifetime of 1.5fs). 
             Reference experimental values (obtained from data in Refs.~\onlinecite{rasigni_josa77,mathewson_philmag72} are represented with red lines. 
            }
       \label{fig:lithium_optical}
    \end{center}
 \end{figure}

As in previous cases we have calculated the optical response of the system by applying a $\delta$-electric
pulse in the z-direction and following the induced current in the system for 20 fs.
At difference with previous examples, however, the minimum densification radius, $\delta r_d=8.0$ \AA~ is 
not enough to obtain converged results and we have tried the values $\delta r_d=$8.0, 10.0, 12.0 \AA.
While the differences found among these results are not extraordinary, we observe some shifts and a better 
resolution of the real-part of the optical conductivity, $\sigma_1$, as we increase the range of the
density matrix (see Fig.~\ref{fig:lithium_optical}).

As with previous examples, we start comparing the experimental data\cite{rasigni_josa77,mathewson_philmag72} 
with the results from pertubation theory in {\sc siesta}. 
We see that this simulation captures an onset of optical absorption around $\hbar \omega \approx$3.5 eV
while the experimental result is somewhat lower and the measured conductivity smaller than the predicted one
by a factor 3.
The observed band is quite dispersive and almost featureless, just showing a second, broad peak at $\hbar \omega\approx$ 8.0 eV
that roughly follows the measured envelope. 
However, at low energies ($\hbar\omega<1.0$ eV), while the experimental data shows a quick increase
related to the Drude peak, the perturbative expression, that only captures interband transitions, does not
show any absorption. 

Moving now on to the TD-SPDFT results we see that the Drude peak, related to a finite conductivity at $\hbar\omega=0$ eV,
is present matching the experimental data. 
Moreover, for relatively small values of $\delta r_d$, the position of the first absortion peak is predicted to be significantly 
lower ($\approx$ 1.3 eV) than the measured one while the second is captured by a peak that is much more localized than the 
broad band in the experiments. 
Increasing the range of the density matrix ($\delta r_d$) allows to obtain results that are computationally more reliable and that show
an improved comparison with the real spectrum. 
In particular we see a progressive shift of the lower peak towards higher energies and a broadening of the second peak
observed in the measurements of Rasigni et al.\cite{rasigni_josa77}
Thus, our TD-SPDFT calculations provide with very satisfactory results when compared to experiment, 
going beyond the simple data-fitting of the initial DFT results, particularly in complicated situations 
like the metallic system studied here.

\begin{figure} [b]
    \begin{center}
       \includegraphics[width=1.0\columnwidth]{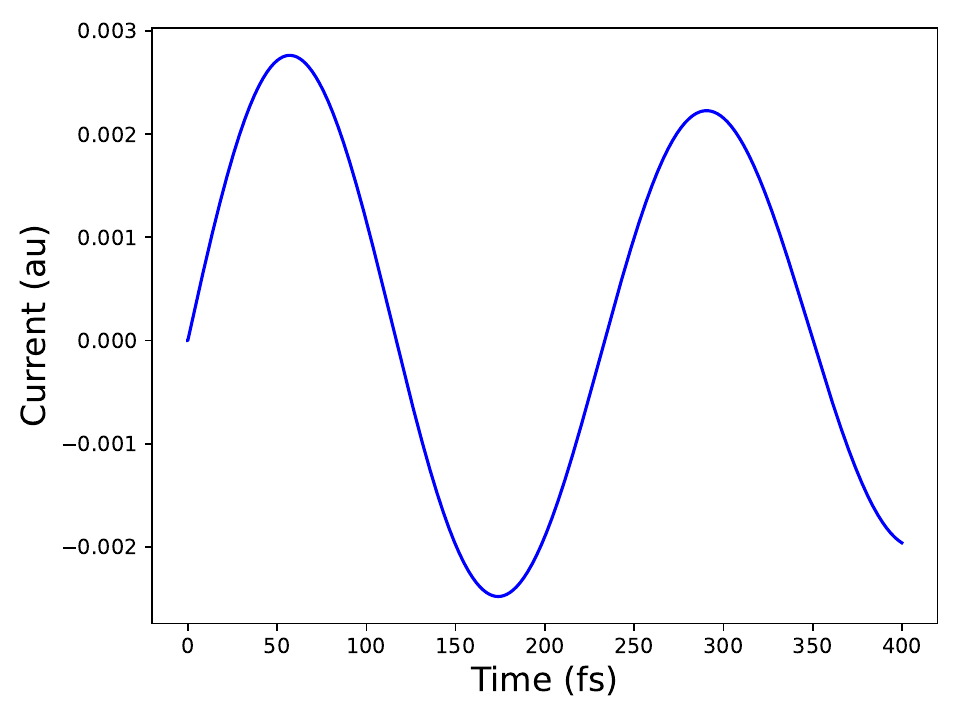}
       \caption{(Color online) Representation of the variation of the current in time in metallic lithium while
                 retaining the geometry frozen to the RAG under a constant electric field of E=10$^{-4}$ au. 
                 The resulting AC current corresponds with the so-called Bloch oscillation.
            }
       \label{fig:lithium_bloch}
    \end{center}
 \end{figure}

To close this section, we will study what happens in our simulation when a constant electric field, $E=10^{-4}$ au, 
is applied on the system, while keeping the geometry frozen to that of the cubic system, to avoid
electron-phonon scattering. 
The expected result is a Bloch oscillation, an alternating current arising from the periodicity of the lattice,
whose frequency is,
\begin{equation}
\omega=\frac{eE_0 a_0}{\hbar} 
\end{equation}
where e is the electron's charge, $E_0$ the electric field intensity and $a_0$ the lattice parameter of the system.
Taking into account that lithium's lattice parameter is 3.45\AA, we set the simulation time to be 400 fs as to observe
more than a full oscillation.
Due to the fact that we are only including the leading terms of Eq.~(\ref{eq:decouple}) in our density propagation, 
we had to use a very small time-step, $\delta t=$10$^{-4}$ fs, to avoid numerical instabilities in the solution.
The result of this simulation is shown in Fig.~\ref{fig:lithium_bloch}, where the current resulting from appliying the
constant electric field is plotted with time. 
It can clearly be seen that the obtained curve oscillates between positive and negative values, creating an AC current
with the characteristic period of the Bloch oscillation. 
While the current implementation of Eq.~(\ref{eq:decouple}) is not very efficient, as higher-orders of the expansion are
needed to increase the time-step to a practical value, the result is a proof-of-concept that TD-SPDFT simulations with 
constant electric fields are possible. 
This fact, combined with models that accurately describe electron-lattice interactions,
will open the possibility of realistically simulate complex phenomena like ohmnic-transport, drift-motion in polarons or 
vibronic effects in optical spectra.

\section{Conclusions}\label{sec:conclusions}

In this work we introduced the time variable into the second-principles scheme presented 
in Ref.~\onlinecite{pgf_prb16}.
This opens up the possibility of studying optical properties in systems containing 
a very large number of atoms ($10^5$) using a very modest computational infrastructure (less than 30 processors).
Given that the method scales linearly with the basis size, it is expected that much larger systems including millions of atoms
could be simulated using massively parallel computers, although this has not been tested yet. 
The key element in this implementation is the use of hamiltonian and density matrices that are fully implemented 
in real-space using periodicity and range-limited interactions, which leads to a sparse representation.
Moreover, we have studied how to deal with electric fields that break the periodicity of the 
lattice. 
We have found that representing the real-space potential, versus choosing a gauge that changed the momentum in time,
was computationally advantageous.
However, when the fields are applied during finite times, as compared to a short pulse, the need to operate
with matrices that are no longer periodic requires ``decoupling'' this part of the hamiltonian from the 
fully periodic part in the spirit of \onlinecite{arrighini_ajp96}. 
While this allows the application of the electric fields in a general way to both metals and insulators,
it hinders the performance of the method, forcing to take very small time steps. 
We expect to improve this aspect in the future by using higher-order expansions.
Finally, we have shown that our real-time SPDFT method is able to significantly improve the predictions
of plain first-order perturbation expressions for the absorption spectra of materials due to: (i) being real-time and thus operating on all orders simultaneously
and (ii) the inclusion of time-dependent electrostatics. 
This results in spectra that are more closely comparable to those obtained with BSE than to simple, perturbation theory in DFT. 
Given that this approach is quite general we suggest that it could also be adapted to standard first-principles packages.

In order to finish we would like to comment on the possible expansions of the presented method.
On one hand the solution involving exponential matrices, Eq.~(\ref{eq:propagation}), could easily be used to express
unitary transformations that could be adapted to solve the SCF problem in large systems.
Similar implementations already exist in what has been ``called the curvy steps for density matrix-based energy minimization'' \cite{helgaker_cpl00,shao_jcp03}.
On the other hand, and in order to fully exploit the method, it is necessary to be able to produce 
models for materials in more effective ways than the ones currently available.
In particular, it is of capital importance to accurately capture electron-lattice and electron-electron interactions.
This would enable us to predict transport properties like relaxation times or polaron difussion with temperature and beyond many 
of the limitations of present methods (like the harmonic approximation, low temperatures, etc). 
This work is now underway.

\begin{acknowledgements}
   We thank M. Moreno and J. A. Aramburu for useful discussions.  PGF
   and JJ acknowledge financial support from Grant No. PID2022-139776NB-C63 funded by
MCIN/AEI/10.13039/501100011033 and by ERDF/EU “A way of making Europe” by the European Union.
T.F.R. acknowledges financial support from Ministerio
de Ciencia, Innovación y Universidades (Grant PRE2019-089054). 
Also, J.Í. acknowledges the financial support
from the Luxembourg National Research Fund through
grant C21/MS/15799044/FERRODYNAMICS.
\end{acknowledgements}

\bibliography{biblio}

\appendix

\section{Time derivatives}
\label{sec:time_der}

In the following we will discuss the formulas that allow to calculate the time derivative 
of an operator expressed using a localized basis.
In particular our basis set consists of WFs which are orthonormal,
\begin{equation}
  \left\langle\chi_{\bm{b}}\right.\left\vert \chi_{\bm{a}}\right\rangle = \delta_{\bm{a}\bm{b}}.
\end{equation}
When we derive the previous expression with time we obtain,
\begin{equation}
  \left\langle\dot{\chi}_{\bm{b}}\right.\left\vert \chi_{\bm{a}}\right\rangle + \left\langle\chi_{\bm{b}}\right.\left\vert \dot{\chi}_{\bm{a}}\right\rangle = 0.
\end{equation}
The equation above allows to show that the time-derivative overlap matrix $\delta S_{ab}$ is antisymmetric,
\begin{equation}
\delta S_{\bm{ab}}=\left\langle\chi_{\bm{a}}\right.\left\vert \dot{\chi}_{\bm{b}}\right\rangle = - \left\langle\dot{\chi}_{\bm{a}}\right.\left\vert \chi_{\bm{b}}\right\rangle = 
\frac{\left\langle\chi_{\bm{a}}\right.\left\vert \dot{\chi}_{\bm{b}}\right\rangle-\left\langle\dot{\chi}_{\bm{a}}\right.\left\vert \chi_{\bm{b}}\right\rangle}{2}=-\delta S_{\bm{ba}}
\end{equation}

If we now consider a generic operator, $\hat{A}$, represented in the WF basis as,
\begin{equation}
\hat{A}=\sum_{\bm{ab}}\left\vert\chi_{\bm{a}}\right\rangle A_{\bm{ab}} \left\langle \chi_{\bm{b}} \right\vert
\end{equation}
we can calculate the time-derivative of $\hat{A}$ in the WF basis,
\begin{widetext}
\begin{align}\label{eq_timeder}
\left\langle\chi_a\right\vert\frac{d\hat{A}}{dt}\left\vert \chi_{\bm{b}} \right\rangle&=
                            \sum_{\bm{cd}}\left(
                                   \left\langle\chi_{\bm{a}}\vert\dot{\chi}_{\bm{c}}\right\rangle A_{\bm{cd}} \left\langle \chi_{\bm{d}} \vert\chi_{\bm{b}} \right\rangle
                                  +\left\langle\chi_a\vert\chi_c\right\rangle \dot{A}_{\bm{cd}} \left\langle \chi_{\bm{d}} \vert\chi_{\bm{b}} \right\rangle
                                  +\left\langle\chi_{\bm{a}}\vert\chi_{\bm{c}}\right\rangle A_{\bm{cd}} \left\langle \dot{\chi}_{\bm{d}} \vert\chi_{\bm{b}} \right\rangle
                             \right)\nonumber\\
                           &=\dot{A}_{\bm{ab}}+\sum_{\bm{c}}\left(
                                   \left\langle\chi_{\bm{a}}\vert\dot{\chi}_{\bm{c}}\right\rangle A_{\bm{cb}}                                   
                                  + A_{\bm{ac}} \left\langle \dot{\chi}_{\bm{c}} \vert\chi_{\bm{b}} \right\rangle
                             \right)\nonumber\\
                           &=\dot{A}_{\bm{ab}}+\sum_{\bm{c}}\left(
                                   \delta S_{\bm{ac}} A_{\bm{cb}}                                   
                                  -A_{\bm{ac}} \delta S_{\bm{cb}}
                             \right)
                            =\dot{A}_{\bm{ab}}+[\delta\hat{S},\hat{A}]_{\bm{ab}}
\end{align}
\end{widetext}

The last term in the above expression is the correction to the matrix elements due to the change of the basis with time. 
This is similar to the Pulay correction\cite{pulay_mp69} applied to the calculation of forces on the atoms of a molecule or solid when using a localized basis. 
We can use the above expression to express the change of the expected value of an operator only due to the change of its basis set.
We will denote this derivate with respect to the basis set with a prime, i.e. $\partial^\prime/\partial^\prime t$,
\begin{widetext}
\begin{align}
\frac{\partial^\prime A_{\bm{ab}}}{\partial^\prime t}&= \left\langle\dot{\chi}_{\bm{a}}\right\vert \hat{A} \left\vert \chi_{\bm{b}} \right\rangle                                    
                                    +\left\langle\chi_a\right\vert \hat{A} \left\vert \dot{\chi}_{\bm{b}}  \right\rangle 
                                   = \sum_{\bm{c}} \left\langle\dot{\chi}_{\bm{a}}\vert\chi_{\bm{c}}\right\rangle \left\langle \chi_{\bm{c}} \right\vert \hat{A} \left\vert \chi_{\bm{b}} \right\rangle
                                    +\sum_{\bm{c}} \left\langle\chi_{\bm{a}}\right\vert \hat{A}\left\vert\chi_{\bm{c}}\right\rangle\left\langle\chi_{\bm{c}} \vert \dot{\chi}_{\bm{b}}  \right\rangle\nonumber\\
                                  &= \sum_{\bm{c}} \left( A_{\bm{ac}}\left\langle\chi_{\bm{c}} \vert \dot{\chi}_{\bm{b}}  \right\rangle
                                                    +\left\langle\dot{\chi}_{\bm{a}}\vert\chi_c\right\rangle A_{\bm{cb}}\right)
                                   =  \sum_{c} \left( A_{\bm{ac}}\delta S_{\bm{cb}}-\delta S_{\bm{ac}}A_{\bm{cb}}\right)
                                   = -[\delta\hat{S},\hat{A}]_{\bm{ab}}\label{eq_partialder}
\end{align}
\end{widetext}

The time derivative of $\hat{A}$ is,~\cite{Cohen-Tannoudji}
\begin{equation}\label{eq_heisenberg}
  \frac{d \hat{A}}{dt}=\frac{i}{\hbar}\left[\hat{h},\hat{A}\right]+\frac{\partial \hat{A}}{\partial t}
\end{equation}
so that substituting Eq.~(\ref{eq_heisenberg}) in Eq.~(\ref{eq_timeder}),
\begin{equation}\label{eq_dotAab}
  \dot{A}_{\bm{ab}}+[\delta\hat{S},\hat{A}]_{\bm{ab}}=\frac{i}{\hbar}\left[\hat{h},\hat{A}\right]_{\bm{ab}}+\left\langle\chi_{\bm{a}}\right\vert\frac{\partial \hat{A}}{\partial t}\left\vert \chi_{\bm{b}} \right\rangle
\end{equation}
and using Eq.~(\ref{eq_partialder}) we get,
\begin{equation}\label{eq_totalder_matrixele}
    \dot{A}_{\bm{ab}} =\frac{i}{\hbar}\left[\hat{h},\hat{A}\right]_{\bm{ab}} + \frac{\partial^\prime A_{\bm{ab}}}{\partial^\prime t} + \left\langle\chi_{\bm{a}}\right\vert\frac{\partial \hat{A}}{\partial t}\left\vert \chi_{\bm{b}} \right\rangle
\end{equation}

Eq.~(\ref{eq_dotAab}) can also be reordered to yield,
\begin{equation}\label{eq_totalder_matrixele2}
  \dot{A}_{\bm{ab}}=\frac{i}{\hbar}\left[\hat{h}+i\hbar\delta\hat{S},\hat{A}\right]_{\bm{ab}}+\left\langle\chi_{\bm{a}}\right\vert\frac{\partial \hat{A}}{\partial t}\left\vert \chi_{\bm{b}} \right\rangle,
\end{equation}
\noindent that can be used to carry out the propagation using a modified Hamiltonian,
\begin{equation}
h^\prime_{\bm{ab}}=\hat{h}_{\bm{ab}}-i\hbar\delta\hat{S}_{\bm{ab}}.
\end{equation}

\end{document}